\documentstyle[psfig,floats,prl,aps,preprint]{revtex}
\begin{document}
\input{psfig.sty}
\draft
\title{Measurement of the 6S$\rightarrow$7S transition polarizablility
 in atomic cesium and an improved test of the standard model}
\author{S. C. Bennett and C. E. Wieman}
\address{JILA, National Institute of Standards and Technology and
  University of Colorado, and\\ Department of Physics,
University of Colorado, Boulder CO 80309-0440}
\date{\today}
\maketitle
\begin{abstract}
  The ratio of the off-diagonal hyperfine amplitude to the tensor
  transition polarizability $(M_{\rm{hf}}/\beta)$ for the
  6S$\rightarrow$7S transition in cesium has been measured. The value
  of $\beta=27.024(43)_{\rm{expt}}(67)_{\rm{theory}}a_{0}^{3}$ is then
  obtained using an accurate semi-empirical value of $M_{\rm{hf}}$.
  This is combined with a previous measurement of parity
  nonconservation in atomic cesium and previous atomic structure
  calculations to determine the value of the weak charge. The
  uncertainties in the atomic structure calculations are updated
  (and reduced) in light of new experimental tests.  The result
  $Q_{W}=-72.06(28)_{\rm{expt}}(34)_{\rm{theory}}$ differs from the
  prediction of the standard model of elementary particle physics by
  2.5$\sigma$.
\end{abstract}
\section{Introduction}
Electroweak experiments have now reached high precision in testing the
standard model and in searching for new physics beyond
it~\cite{Peskin:summary,Langacker}. These experiments include
measurements of parity nonconservation (PNC) in atoms as first
proposed in Ref.~\cite{Bouchiatboth}. Atomic PNC measurements are
uniquely sensitive to a variety of new physics, such as the existence
of additional $Z$ bosons, because of the different energy scale and
because they probe a different set of model-independent electron-quark
coupling constants than those measured by high-energy
experiments~\cite{Langacker}.  The most precise atomic PNC
experiment~\cite{Wood97} examines the mixing of $S$ and $P$ states in
atomic cesium.  Specifically, it compares the mixing due to the PNC
neutral weak current interaction to the $S$-$P$ mixing caused by an
applied electric field (``Stark mixing'').  In previous
work~\cite{Wood97}, this measurement was combined with theoretical
calculations of the structure of the cesium atom to obtain the weak
charge $Q_{W}$, which characterizes the strength of the neutral weak
interaction and can be compared to the value predicted by the standard
model.  The atomic structure calculations were used to obtain two
pieces of information: the amount of Stark mixing and the relevant PNC
electronic matrix elements.  The 1.2\% uncertainty in the
determination of $Q_{W}$ was dominated by the uncertainties in those
two calculated quantities. In this paper we report a reduced
uncertainty in $Q_{W}$ that is obtained by 1) measuring the Stark
mixing, and 2) incorporating new experimental data into the evaluation
of the uncertainty in the calculation of the PNC matrix elements.
These new data indicate that the calculations are more accurate than
was indicated by the less precise (and in some cases incorrect) data
available at the time the calculations were published.

\section{Theory}
The $6S$ ground state and $7S$ excited state of atomic cesium both have
two hyperfine levels: $F=3$ and $F=4$. In the presence of a dc electric
field $\vec{E}$, a magnetic field, and a standing-wave laser
field with propagation vector $\vec{k}$ and polarization
$\vec{\epsilon}$, the $\Delta F=\pm1$ $6S\rightarrow7S$ amplitudes,
used in both Ref.~\cite{Wood97} and the present work, are given
by~\cite{Gilbert86}
\begin{equation}
A_{6S\rightarrow7S}=[i\beta(\vec{E}\times\vec{\epsilon}\:)+
  M1(\vec{k}\times\vec{\epsilon})+
  E1_{\rm{PNC}}\vec{\epsilon}\,]\cdot\langle
  F'm_{F}'\mid\vec{\sigma}\mid Fm_{F}\rangle,
\end{equation}
where $M1=M\pm M_{\rm{hf}}\delta_{FF'\pm1}$ is the magnetic dipole
amplitude ($M$ is from relativistic and spin-orbit effects,
$M_{\rm{hf}}$ is from the off-diagonal hyperfine interaction), and
$\vec{\sigma}$ is the Pauli spin matrix.  The tensor transition
polarizability~\cite{Bouchiatboth} $\beta$ characterizes the size of the
Stark mixing-induced electric dipole amplitude, and $E1_{\rm{PNC}}$ is
the PNC matrix element given by
\begin{equation}
E1_{\rm{PNC}}\equiv \overline{\langle 7S \mid} \bbox{D} \overline{\mid
  6S\rangle}=\frac{Q_{W}}{N} k_{\rm{PNC}}.
\end{equation}
Here, $\overline{\mid\! nS\rangle}$ is an $\mid\! nS
\rangle$ state into which the PNC Hamiltonian has mixed a small
amount of $\mid\! nP\rangle$ states, $\bbox{D}$ is the electric dipole
operator, $N$ is the number of neutrons, and $k_{\rm{PNC}}$ is the
calculation of the sum of relevant matrix elements between $S$ and $P$
states given by
\begin{equation}\label{kpnc}
k_{\rm{PNC}}=\frac{N}{Q_{W}}\sum_{n}\left(\frac{\langle 7S\mid
\vec{D} \mid nP\rangle\langle nP\mid H_{\rm{PNC}}\mid
6S\rangle}{E_{6S}-E_{nP}}
+\frac{\langle 7S\mid H_{\rm{PNC}} \mid nP\rangle\langle nP\mid
    \vec{D}\mid 6S\rangle}{E_{7S}-E_{nP}}\right).
\end{equation}
Since $H_{\rm{PNC}}=G_{F} \gamma^{5} Q_{W}\rho_{N}(r)/\sqrt{8}$, each
of the terms in Eq.~(\ref{kpnc}) is the product of a dipole matrix
element times a $\gamma^{5}$ matrix element evaluated in the nucleus.
Ninety-eight percent of the sum comes from the $6P_{1/2}$ and
$7P_{1/2}$ states~\cite{Blundellother}.

In Ref.~\cite{Wood97}, $\rm{Im}(E1_{\rm{PNC}})/\beta$ is measured.
The value $Q_{W}$ is obtained by multiplying this ratio by $\beta
N/k_{\rm{PNC}}$.  This paper concerns the improved determination of
$\beta$ and $k_{\rm{PNC}}$, and thus $Q_{W}$.

To determine $\beta$, we measure $M_{\rm{hf}}/\beta$ and take
advantage of the fact that $M_{\rm{hf}}$ can be accurately determined
semi-empirically~\cite{Bouchiat:E2expt}.  The amplitude $M_{\rm{hf}}$
is due to the hyperfine interaction and thus can be expressed in terms
of well-measured hyperfine splittings. In this experiment we observe
the $6S\rightarrow7S$ rate driven with a standing-wave laser beam with
polarization $\vec{\epsilon}=\epsilon\hat{z}$ and a field geometry
($E$ along $\hat{x}$) such that the transition rate is
\begin{equation}\label{betarate}
\mid A_{\rm{6S\rightarrow 7S}}\mid^{2}=\beta^{2}E^{2}\epsilon^{2}
+(M\pm M_{\rm{hf}}\delta_{FF'\pm1})^{2}\epsilon^{2},
\end{equation}
where small interference terms have been omitted.  The $\beta$-PNC and
$M1$-PNC interference terms are negligible and the $\beta$-$M1$
interference terms cancel almost identically $(<10^{-6})$ because of
their $\hat{k}$ dependence and the standing-wave geometry of the
experiment. We determine $M_{\rm{hf}}/\beta$ by measuring the total
rate on the two $\Delta F=\pm 1$ hyperfine transitions with large $E$,
where $\mid\!  A_{\rm{6S\rightarrow7S}}\!  \mid^{2}\approx
\beta^{2}E^{2}$, and with $E=0$, where $\mid\!
A_{\rm{6S\rightarrow7S}}\!\mid^{2} \approx(M\pm
M_{\rm{hf}}\delta_{FF'\pm1})^{2}$. We combine the ratios of the high
and low $E$ rates on both transitions to determine
$M_{\rm{hf}}/\beta$.

A complication arises because the locations of the antinodes of the
oscillating electric ($\varepsilon_{ac}$) and magnetic ($b_{ac}$)
fields are separated by $\lambda/4$ in the standing wave.  Because of
this separation, photoionization (which is driven by
$\varepsilon_{ac}$) is larger for $7S$ atoms excited by
$\varepsilon_{ac}$ (E1 atoms) than it is for $7S$ atoms excited by
$b_{ac}$ ($M1$ atoms). The result is that the detection efficiency for
$E1$ excitations is slightly smaller ($\sim$ 1\% for typical
intensities) than for $M1$ excitations.  This difference gives a
potential systematic error that is intensity dependent.  The ratios of
the signals, measured at a laser intensity $I$, for the $\Delta F=+1$
and $\Delta F=-1$ transitions, respectively, are then
\begin{mathletters}\label{ratios}
\begin{equation}
R^{3\rightarrow 4}_{I}\equiv(\frac{M-M_{\rm{hf}}}{\beta E})^{2}(1+\eta
I),\rm{ and}
\end{equation}
\begin{equation}
R^{4\rightarrow 3}_{I}\equiv(\frac{M+M_{\rm{hf}}}{\beta E})^{2}(1+\eta I),
\end{equation}
\end{mathletters}
where $\eta$ is a parameter that describes the difference in
photoionization fraction.

\section{Experiment}\label{betameas}
The apparatus used in the present experiment is very similar to that
in Refs.~\cite{Wood97,dcStark}. A collimated beam of cesium is
optically pumped into either the $F=3$ or $F=4$ hyperfine level of the
$6S_{1/2}$ ground state.  The beam of atoms then travels roughly along
the $\hat{z}$ axis into a region with mutually orthogonal dc electric
(along $\hat{x})$ and magnetic (along $\hat{z})$ fields and intersects
a 540~nm standing-wave laser field (along $\hat{y})$ at right angles.
The laser field is produced by a tunable dye laser that is
frequency-locked to a finesse $\simeq 10^{5}$ Fabry-Perot etalon.  The
etalon is, in turn, locked to a stable reference cavity.  The light
going to the reference cavity is double-passed through an
acousto-optic modulator (AOM), so the frequency of the laser light
interacting with the atomic beam is
$\nu_{\rm{laser}}=\nu_{\rm{reference}}-2\nu_{\rm{AOM}}$.  Thus, we can
change the frequency of the dye laser in a very controlled manner by
changing the frequency of the AOM.  The dye laser drives the
$6S\rightarrow7S$ transition.  Approximately half of the atoms excited
to the $7S$ state relax to the previously depleted hyperfine ground
state ($F=3$ or $F=4$). Further downstream, the atoms in the
repopulated hyperfine level scatter photons from a diode laser probe
beam tuned to an appropriate $6S_{1/2}$-$6P_{3/2}$ cycling transition.
We collect the scattered photons on a large-area photodiode, and its
photocurrent is proportional to the number of atoms making the
6S$\rightarrow$7S transition.

To measure the ratio $R^{3\rightarrow 4}_{I}$ (or $R^{4\rightarrow
  3}_{I}$), we scan the laser over the $6S\rightarrow7S$ $\Delta F=+1$
(or $\Delta F=-1$) transition in 0.3~MHz steps.  After each step we
integrate the photocurrent for 16.67~ms and store that data point on
disk.  We alternate between scans with $E=707.63(68)$~V/cm and $E=0$~V/cm.

There is a 540~nm-laser-frequency-independent background signal from atoms in
the wrong hyperfine state that is $\sim 100$ times larger than the
desired $M1$ signal for $E=0$~V/cm. We measure this background before
and after each data point by detuning the laser $\sim$~50~MHz from
line center and measuring the photocurrent.  These background points
are measured alternately above and below the line center to cancel any
linear frequency dependence of the background.  We subtract the
average background from the data points to leave only the contribution
from atoms making the 6S$\rightarrow$7S transition.  The sum of all
the data points (the area under the spectral line) is proportional to
the total transition rate.

We looked for but did not observe any frequency dependence to the
background.  Also, all likely mechanisms, such as molecular
transitions or light scattering off the mirrors, should have very
broad spectral features, and hence will be eliminated by the
background subtraction. The uncertainty in our results due to possible
frequency dependent backgrounds is less than 0.05\%.

Sample background-subtracted scans are shown in Fig.~\ref{datasamp}.
The two line shapes are asymmetric and slightly offset from one
another because of their differing sensitivity to ac Stark shifts as
discussed in Ref.~\cite{lineshape}.  The different line shapes do not
affect our measurement of the total transition rate because the atoms'
total transition amplitude is unchanged, even though the resonant
frequency of each atom is shifted according to the local
$\varepsilon_{\rm{ac}}$ field.  Therefore, by integrating the areas
under the entire broadened lines we can determine the desired relative
ratios $R^{3\rightarrow 4}_{I}$ and $R^{4\rightarrow 3}_{I}$.

\section{Results}
The detection efficiency and signal-to-noise ratio are significantly
higher for $R^{3\rightarrow 4}$; we measure that ratio at five
different intensities from 0.6~kW to 2.8~kW and determine
$\eta$ to 1.5 parts in $10^{3}$ using a least squares fit. We
find the ratios $R^{3\rightarrow 4}_{0}=2.4636(8)\times 10^{-3}$ and
$R^{4\rightarrow 3}_{0}=1.1357(6)\times 10^{-3}$ where the two
uncertainties have a common contribution from the extrapolation to
zero intensity.  Combining these results using Eq.~(\ref{ratios}) we
find $M_{\rm{hf}}/\beta=-5.6195(91)\rm{V/cm}$.~\cite{E2comment}. From
Ref.~\cite{Bouchiat:E2expt} we take
$M_{\rm{hf}}=-151.86(38)\rm{(V/cm)} a_{0}^{3}$, which is based on
measured hyperfine splittings with a $0.3 \pm 0.3\%$ theory correction
due to many body effects. This gives
\begin{equation}
\beta=27.024(43)_{\rm{expt}}(67)_{\rm{theory}}a_{0}^{3}.
\end{equation}
This value is in excellent agreement with the semi-empirical
values $\beta=27.17(35)a_{0}^{3}$~\cite{Bouchiat:E2expt} and
$\beta=27.15(13)a_{0}^{3}$~\cite{Dzuba97}, and the calculated value
$\beta=27.00a_{0}^{3}$~\cite{BlundellPNC}.

Using our measured values for $\beta$ and Im($E1_{\rm{PNC}})/\beta$,
and the calculated value of $k_{\rm{PNC}}$, we can now extract
$Q_{W}$. The key issue is the uncertainty in the value of
$k_{\rm{PNC}}$.  The authors of
Refs.~\cite{Blundellother,BlundellPNC,DzubaPNC,Dzubaother} discuss
this issue at considerable length. Here we only summarize the
conclusion of both groups that the most reliable measure is to use the
same {\it ab initio} calculations of the electronic structure that are
used to find $k_{\rm{PNC}}$ to calculate dipole matrix elements and
hyperfine splittings for the $6S_{1/2}$, $7S_{1/2}$, $6P_{1/2}$, and
$7P_{1/2}$ states. The differences between these calculated values and
the experimental determinations provide a reliable quantitative
indication of the uncertainties in the calculations of $k_{\rm{PNC}}$.
The authors considered how well these errors in the hyperfine
splittings and dipole matrix elements reflect errors in $k_{\rm{PNC}}$
by rescaling their calculations in a variety of ways and comparing the
relative sensitivities of the different quantities.  They found that
$k_{\rm{PNC}}$ has comparable or smaller sensitivity than the other
quantities~\cite{kpnccomment}.  From comparing calculated and measured
quantities, both groups arrived at uncertainties of about 1\% for
their value of $k_{\rm{PNC}}$.  Since the time that
Refs.~\cite{Blundellother,BlundellPNC,DzubaPNC,Dzubaother} were
published, there have been a number of new and more precise
measurements of the quantities of interest.  In all cases, the new
measurements show better agreement with the calculations than earlier
measurements and also show that the largest previous disagreements
were likely due to experimental errors.

In Table~\ref{comptable} we have collected the results of the most
precise measurements of relevant quantities in cesium.  We list the
quantities measured, the primary aspect of the electronic wave
functions that is being tested in each comparison, and the difference
between theory and experiment. Particularly notable are the top three
lines of the table, which show that the agreement has dramatically
improved from the 1-2\% disagreements of the older experiments. In
addition to the data in this table, there have been new experiments
that revealed errors in earlier lifetime measurements in sodium and
lithium.  These new data eliminate what had appeared to be troubling
1\% errors in equivalent calculations for those atoms.

The standard deviation of the fractional differences between theory
and experiment in Table~\ref{comptable} is $4.0\times10^{-3}$. We
believe this to be the most valid number to use to represent the 68\%
confidence level for $k_{\rm{PNC}}$.  Using the average of
$k_{\rm{PNC}}=0.905\times 10^{-11}iea_{0}$~\cite{BlundellPNC} and
$k_{\rm{PNC}}=0.908\times 10^{-11}iea_{0}$~\cite{DzubaPNC} this gives
a value of $k_{\rm{PNC}}=0.9065(36)\times 10^{-11}iea_{0}$

When combined with our new value for $\beta$ and the experimental PNC
measurement this gives
\begin{equation}
  Q_{W}=-72.06(28)_{\rm{expt}}(34)_{\rm{theory}}.
\end{equation}
The standard model value including radiative corrections is
$Q_{W}=-73.20(13)$~\cite{marciano}. Adding the uncertainties in
quadrature, these values differ by 2.5 $\sigma$.

Assuming that this difference is not due to an experimental error or a
statistical fluctuation, it suggests several possibilities. The first
possibility is that the calculated value of the $\gamma^{5}$ matrix
element is in error by the requisite 1.58\%. In light of Table I, such
an error would require a wave function with a somewhat peculiar and
insidious shape. Although none of the measured quantities depends on
the shape of the wave function in a manner identical to that of
$\gamma^{5}$, the different comparisons in Table I do probe the value
of the wave function in all regions: short, intermediate, and long
distances.  The largest single difference of the 16 comparisons is
only $0.79\%$, and the standard deviation is only $0.40\%$.  The
second possibility is that there are contributions or corrections to
atomic PNC within the standard model that have been overlooked. We see
no justification for either of these two possibilities, but they clearly
need to be explored further.  The first offers a formidable but not
overwhelming challenge to both theoretical and experimental atomic
physicists.

The final possibility is that this discrepancy is indicating the
presence of some new physics not contained in the standard model.
Physics that would be characterized by the S
parameter~\cite{Peskin:Sparam} is not a likely candidate because the
size of the contribution needed [$S= -1.4(6)$] would be in conflict
with other data~\cite{Peskin:summary}.  However, there are other types
of new physics, such as an additional Z boson, that would be consistent
all other current data.

\section{Acknowledgements}
We are happy to acknowledge support from the NSF, assistance in the
experiments by J. L. Roberts, and valuable discussions with V. V.
Flambaum, P. G. H. Sandars, C. E. Tanner, and W. R. Johnson.  V. A.
Dzuba graciously sent us his tabulation of calculated matrix elements
as well as providing other valuable comments.

\begin{table}[p]
  \caption{Fractional differences ($\times 10^{3}$) between measured
  and calculated values of quantities relevant for testing PNC
  calculations in atomic cesium.  We only list the most precise
  experiments.  The second column lists the most relevant aspects of
  the wavefunctions that are being tested.  $\langle
  1/r^{3}\rangle_{nP}$ is the average of $1/r^{3}$ over the
  wavefunction of the electronic state $nP$. Where the experiment has
  improved or changed significantly since the publication of
  Ref.~\protect\cite{BlundellPNC}, the difference from the old
  experiment is listed in brackets.}\label{comptable}
\begin{tabular}{ccccc}
  Quantity & Calculation & \multicolumn{2}{c}{Difference $(\times 10^{3})$}& \\
  measured & tested &Dzuba, {\it et
  al.}\tablenotemark[1]$^{,}$\tablenotemark[2] & Blundell {\it et
  al.}\tablenotemark[3]& $\sigma_{Expt}$\\\hline
  6S$\rightarrow$7S dc Stark shift\tablenotemark[4]& $\langle
  7P\mid\mid \bbox{D}\mid\mid 7S \rangle$ & $-$3.4[19] &
  $-$0.7[22] & 1.0[4]\\
  $6P_{1/2}$ lifetime\tablenotemark[5] & $\langle 6S\mid\mid \bbox{D}
  \mid\mid
  6P_{1/2}\rangle$& $-$4.2[$-$8]&4.3[1]&1.0[43]\\
  $6P_{3/2}$ lifetime\tablenotemark[5] & $\langle 6S\mid\mid \bbox{D}
  \mid\mid
  6P_{3/2}\rangle$& $-$2.6[$-$41]&7.9[$-$31]&2.3[22]\\
  $\alpha$\tablenotemark[6]&
  $\langle 7S\mid\mid\bbox{D}\mid\mid 6P_{1/2}\rangle$, and&   &    & \\
                           &
  $\langle 7S\mid\mid\bbox{D}\mid\mid 6P_{3/2}\rangle$& $-$ &$-$1.4
  & 3.2\\
  $\beta$\tablenotemark[7]  & same as $\alpha$ &$-$ &$-$0.8&3.0\\
  6S HFS\tablenotemark[8]&$\psi_{6S}(r=0)$  & 1.8&$-$3.1&$-$\\
  7S HFS\tablenotemark[9]&$\psi_{7S}(r=0)$  & $-$6.0&$-$3.4&0.2\\
  6P$_{1/2}$ HFS\tablenotemark[10] & $\langle 1/r^{3}\rangle_{6P}$
                                                      & $-$6.1&2.6&0.2\\
  7P$_{1/2}$ HFS\tablenotemark[11] & $\langle 1/r^{3}\rangle_{7P}$
                                             & $-$7.1&$-$1.5&0.5
\end{tabular}
\tablenotemark[1]The value for $k_{\rm{PNC}}$ of Dzuba, {\it et al.}
  is obtained using ``energy rescaling'' so we have used the
  corresponding "rescaled" values in the table for consistency.
  Blundell {\it et al.} do not rescale $k_{\rm{PNC}}$ and so we use
  their pure {\it ab inito} values in the table.
\tablenotemark[2]~{Refs.~\cite{DzubaPNC,Dzubaother}};
\tablenotemark[3]~{Refs.~\cite{BlundellPNC,Blundellother};}
\tablenotemark[4]~{Ref.~\cite{dcStark};}
\tablenotemark[5]~{Ref.~\cite{6Slifetime};}
\tablenotemark[6]~{Using present work's value of $\beta$ and
  $\alpha/\beta$ from Ref.~\cite{alphatobeta};}
\tablenotemark[7]~{Present work;} \tablenotemark[8]~{Defined;}
\tablenotemark[9]~{Ref.~\cite{GilbertHFS};}
\tablenotemark[10]~{Ref.~\cite{Tanner:new};}
\tablenotemark[11]~{Ref.~\cite{ArimondoHFS};}
\end{table}

\begin{figure}[p]
\begin{center}
\end{center}
%this figure should be ``betasamp.eps''
\caption{Sample data comparing scans with and without an applied
  electric field.  Open circles are with $E=707$~V/cm, and the scale
  on the right. Closed circles are with $E=0$~V/cm and the scale on
  the left.  The two lines are offset from one another and have
  different widths because of the different sensitivities to ac Stark
  shifts for the $M1$ and $E1$ transitions.}
  \label{datasamp}
\end{figure}

\begin{thebibliography}{99}
\bibitem{Peskin:summary}A review of the most recent high-energy
  experiments can be found in M. E. Peskin, Science {\bf 281}, 1153
  (1998).
\bibitem{Langacker}P. Langacker, M. Luo, and A. K. Mann,
  Rev. Mod. Phys. {\bf 64}, 87 (1992).
\bibitem{Bouchiatboth}M. A. Bouchiat and C. Bouchiat, J. Phys. (Paris).
  {\bf 35}, 899 (1974), {\it ibid.}, {\bf 36}, 493 (1975).
\bibitem{Wood97}C. S. Wood, {\it et al.}, Science {\bf 275}, 1759
  (1997).
\bibitem{Gilbert86}S. L. Gilbert and C. E. Wieman, Phys. Rev. A {\bf
  34}, 792 (1986).
\bibitem{Blundellother}S. A. Blundell, J. Sapirstein, and
  W. R. Johnson, Phys. Rev. A {\bf 43}, 3407 (1991); S. A. Blundell,
  W. R. Johnson, and J. Sapirstein, {\it ibid.}, {\bf 42} 3751 (1990);
  Phys. Rev. Lett. {\bf 65}, 1411 (1990).
\bibitem{Bouchiat:E2expt}M. A. Bouchiat and J. Gu\'{e}na, J. Phys.
  (France) {\bf 49}, 2037 (1988).
\bibitem{dcStark}S. C. Bennett, J. L. Roberts, and C. E. Wieman,
  Phys. Rev. A {\bf 59}, R16 (1999).
\bibitem{lineshape}C. E. Wieman {\it et al.}, Phys. Rev. Lett. {\bf
  58}, 1738 (1987).
\bibitem{E2comment}We have also included a $-2.3\times 10^{-3}$
  correction to $M_{\rm hf}/\beta$ to account for the effects of the
  electric quadrupole interaction, $E2$, discussed in
  Refs.~\cite{Bouchiat:E2expt,Bouchiat:E2theo}. This correction
  depends on the differences in the $m_{F}$ sublevel populations
  induced by the optical pumping into a single hyperfine state. The
  present work determines the value $M_{\rm{hf}}/M=-0.1906(3)$. By comparing
  this to the result in Ref.~\cite{Gilbert86:2}, we find
  $E2/M_{\rm{hf}}=53(3)\times 10^{-3}.$ These agree with the less
  precise values $M_{\rm{hf}}/M=-0.1886(17)$, and
  $E2/M_{\rm{hf}}=42(13)\times 10^{-3}$ found in
  Ref.~\cite{Bouchiat:E2expt}. A detailed discussion of these issues
  can be found in Ref.~\cite{Bennett:thesis}
\bibitem{Dzuba97}V. A. Dzuba, V. V. Flambaum, and O. P. Sushkov,
  Phys. Rev. A {\bf 56}, R4357 (1997).
\bibitem{BlundellPNC}S. A. Blundell, J. Sapirstein, and W. R. Johnson,
  Phys. Rev. D {\bf 45}, 1602 (1992).
\bibitem{DzubaPNC}V. A. Dzuba, V. V. Flambaum, and O. P. Sushkov,
  Phys.  Lett. A {\bf 141}, 147 (1989).
\bibitem{Dzubaother} V. A. Dzuba, V. V. Flambaum, and O. P. Sushkov,
  Phys. Lett. A {\bf 140}, 493 (1989); V. A. Dzuba, {\it et al.}, {\it
  ibid}, {\bf 142}, 373 (1989).
\bibitem{kpnccomment}This seems reasonable since the individual
dipole and hyperfine matrix elements will likely be more sensitive to
local errors in the wave function (at long and short ranges
respectively) than will be the products of matrix elements that enter
into Eq.~(\ref{kpnc}).
\bibitem{marciano}W. J. Marciano and J. L. Rosner, Phys. Rev. Lett. {\bf
    65}, 2963 (1990); {\bf 68}, 898(E) (1992).
\bibitem{Peskin:Sparam}M. E. Peskin and T. Takeuchi,
  Phys. Rev. Lett. {\bf 65}, 964 (1990).
\bibitem{6Slifetime}R. J. Rafac, {\it et al.}, to be published.
\bibitem{alphatobeta}D. Cho, {\it et al.}, Phys. Rev. A {\bf 55}, 1007
  (1997).
\bibitem{GilbertHFS}S. L. Gilbert, R. N. Watts, and C. E. Wieman,
  Phys. Rev. A. {\bf 27}, 581 (1983).
\bibitem{Tanner:new}R. J. Rafac and C. E. Tanner, Phys. Rev. A {\bf
  56}, 1027 (1997)
\bibitem{ArimondoHFS}E. Arimondo, M. Inguscio, and P. Violino,
  Rev. Mod. Phys. {\bf 49}, 31 (1977).
\bibitem{Bouchiat:E2theo}M. A. Bouchiat and L. Pottier, J. Phys.
  (France) {\bf 49}, 1851 (1988).
\bibitem{Gilbert86:2}S. L. Gilbert {\it et. al.}, Phys. Rev. A {\bf
  34}, 3509 (1986).
\bibitem{Bennett:thesis}S. C. Bennett, Ph.D. thesis, University of
  Colorado,1998, unpublished.
\end{thebibliography}
\end{document}